\documentclass[amsfonts, amssymb, amsmath, twocolumn, showkeys, nofootinbib,superscriptaddress, twoside, aps, prx, titles, colorlinks, citecolor=blue, linkcolor=blue, urlcolor=blue]{revtex4-2}
\usepackage[english]{babel}
\usepackage[utf8]{inputenc}
\usepackage[colorinlistoftodos, color=green!40, prependcaption]{todonotes}
\usepackage{tabularx}
\usepackage{amsthm}
\usepackage{mathtools}
\usepackage{physics}
\usepackage{xcolor, color}
\usepackage{graphicx}
\usepackage{adjustbox}
\usepackage{placeins}
\usepackage[T1]{fontenc}
\usepackage{lipsum}
\usepackage{csquotes}
\usepackage{svg}
\usepackage[pdftex, pdftitle={Article}, pdfauthor={Author}]{hyperref} 
\usepackage{soul}
\usepackage{multirow}

\definecolor{darkgreen}{rgb}{0.2,0.7,0}

\begin{document}

\title{Spectroscopic measurement of the Casimir-Polder force in the intermediate regime}

\author{K. Ton} 
    \affiliation{Department of Physics, University of California San Diego, California 92093, USA}

\author{G. Kestler} 
    \affiliation{Department of Physics, University of California San Diego, California 92093, USA}

\author{D. A. Steck} 
    \affiliation{Department of Physics and Oregon Center for Optical, Molecular, \& Quantum Sciences, University of Oregon, Eugene, Oregon 97403}
    
\author{J. T. Barreiro} 
    \affiliation{Department of Physics, University of California San Diego, California 92093, USA}
    
\begin{abstract}

The Casimir-Polder (CP) effect---the force between a neutral atom and an uncharged conducting plate in empty space---is an intriguing consequence of quantum vacuum fluctuations. The typically attractive CP potential crosses over from a scaling of $z^{-3}$ at short separations to $z^{-4}$ at long distances, where retardation effects due to the finite speed of light become important. At intermediate distances, where the atom--surface separation is of the order of the wavelength of the dominant atomic transition, experiments have so far relied on indirect methods, such as diffraction or quantum reflection, to observe the CP effect. 
Here, we directly reveal the CP force between strontium atoms and a dielectric surface via the induced shifts in the atomic energy levels in the intermediate regime. We spectroscopically probe the CP-induced kHz-frequency shift of ultracold atoms confined by a magic-wavelength optical lattice at 189(2)~nm from the surface---on the scale of the dominant 461-nm transition.  
Our measurements agree well with QED calculations and differ from the short-range approximation, while excluding the long-distance one. This paves the way for studying the CP effect across various surface properties and geometries, as well as exploring the tensor nature of the atom-surface potential---all important for the development of hybrid atomic optical-magnetic quantum devices.

\end{abstract}

\maketitle

The rapid advancement of novel hybrid quantum devices that trap atoms near the surfaces of bulk materials has made the accurate estimation of the Casimir-Polder potential increasingly important~\cite{Vetsch2010,Goban2012,Goban2015,Patterson2018,Kestler2023,Zhou2024}. This estimation is often performed through theoretical calculations~\cite{Le2007,Le2022} or numerical electromagnetic simulations~\cite{Stern_2011,Hung_2013, Reid_2009, Johnson_2011,Reid_2013}. A lack of precise knowledge of the surface potential can lead to many trial-and-error attempts before achieving an optimal design. Despite this challenge, interest in atom-trapping methods that rely on the Casimir-Polder potential continues to grow~\cite{Chang2014, Gonzalez-Tudela2015}. These platforms usually confine atoms to within a few hundred nanometers of the surface, placing them in the intermediate regime where precise spectroscopic measurements of the Casimir-Polder potential are currently lacking.

Since the theoretical description of the Casimir-Polder (CP) force~\cite{Casimir1948}, several experiments have set out to measure the effect. Early attempts used Rydberg atoms, mainly investigating the near-field regime, due to the longer dominant atomic transition wavelengths~\cite{Raskin1969,Shih1975,Anderson1988,Sandoghdar1992,Landragin1996}. The near-field surface force, often referred to as the van~der~Waals force, is valid for an atom-surface separation 
$z \ll  \lambda_0$,
where $\lambda_0$ is the dominant transition wavelength of the atom. In the large distance limit 
($z \gg \lambda_0$),
retardation effects become important. With the development of laser cooling, cold atoms were employed to probe the far-field retarded regime. In the retarded regime, the CP potentials were extracted via quantum reflections~\cite{Shimizu2001, Pasquini2004}, atom interferometry~\cite{Marani2000}, and the effect on the mechanical motion of the Bose-Einstein condensate~\cite{Harber2005}. In the intermediate-distance regime 
($z \sim\lambda_0$),
the CP potential is best described by the full quantum electrodynamics (QED) treatment. Some experiments have measured the CP force in this crossover regime through diffraction or reflection experiments~\cite{Sukenik1993,Bender2010,Garcion2021}. Here, the data indicate that the CP potential exhibits neither $z^{-3}$ nor $z^{-4}$ dependence, but is in close agreement with the full QED calculation.   

Strontium atoms have a narrow-linewidth ${}^1S_0-{}^3P_1$ intercombination transition that makes them ideal for high-resolution spectroscopy that resolves the subtle energy-level shifts caused by the CP effect from a dielectric surface. Another advantage of using ${}^{88}$Sr is that its spherically symmetric ground state is insensitive to magnetic-field noise, and the small collisional scattering length reduces frequency shifts arising from interatomic collisions. Then, when a strontium atom is brought close to a surface, the differences in polarizability cause the ground state ${}^1S_0$ and the excited state ${}^3P_1$ to experience different shifts from the CP potential (Appendix~\ref{app:Theory}). Consequently, the transition frequency is shifted as a function of the atom-surface distance. 

In this study, we report the first direct spectroscopic measurement of the Casimir-Polder potential in the intermediate-distance regime. This experimental measurement agrees well with the theoretical QED calculation result, proving its accuracy. More importantly, our energy shift measurement achieves an error of a few kilohertz, which is at least an order of magnitude higher precision than previous measurements in the intermediate-range \cite{Bender2010}. This work opens up a new paradigm for precisely probing the Casimir-Polder potential. 

Spectroscopically measuring the Casimir-Polder shift in the intermediate regime requires (i) cold neutral atoms trapped from the CP surface in the range of the dominant transition wavelength of strontium at 461~nm [Fig.~\ref{fig:experiment}(a)], (ii) a narrow-linewidth probing transition that can resolve the CP induced shift, and (iii) sufficiently high signal-to-noise ratio (SNR) of the measurement technique. Our experiment utilizes bosonic $^{88}$Sr atoms to achieve each of these requirements. 

\section*{Trapping strontium near dielectric surface}

\begin{figure}[t!]
     \includegraphics[width=\columnwidth]{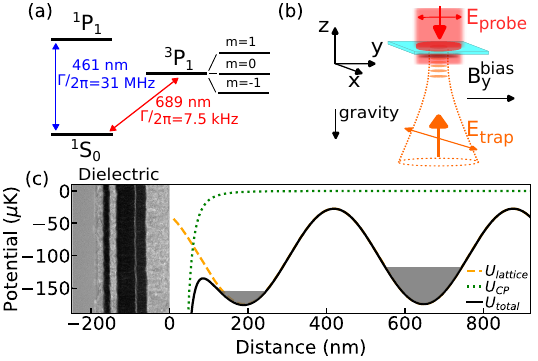}
     \caption{\textbf{Experimental setup: atomic and surface properties.} (a) Relevant energy levels and transitions for ${}^{88}$Sr used in the experiment. (b) The 914-nm optical lattice beam is launched from below and focused onto the CP test surface. The probe laser propagates downward to illuminate the lattice-trapped atoms. A bias magnetic field, ${B}^{\text{bias}}_{\text{y}}$, is applied parallel to the probe electric field, $\mathbf{E}_{\text{probe}}$, while being orthogonal to the lattice beam polarization, $\mathbf{E}_{\text{trap}}$. (c) Calculated trapping potential versus distance from the test surface to strontium atoms (grey filling) in the ground state, given the measured lattice parameters. The total trapping potential (solid black) is the sum of the CP potential (dotted green) and the optical lattice potential (dashed orange). The surface's optical thin-film coating layers, as observed under transmission electron microscope (TEM) imaging, are shown. We calculate that the first lattice site is located at $189$~nm from the surface and has a trap depth of $37\ \mathrm{\mu K}$.}
   \label{fig:experiment}
\end{figure}

We address (i) by first cooling ${}^{88}$Sr atoms to around $1\,\mu$K using standard laser-cooling techniques and, second, by trapping these atoms 189(2)~nm from a dielectric surface via a 914~nm magic-wavelength optical lattice~\cite{Ido2003}. In the first step, we prepare a cloud of $5\times10^4$ atoms at around $1\ \mathrm{\mu K}$ with a disk shape approximately $70\ \mathrm{\mu m}$ thick and $250\ \mathrm{\mu m}$ in diameter as described in Appendix~\ref{app:Prep}. For the second step, the magic-wavelength trapping field offers zero differential ac Stark shift on the ${}^1S_0-{}^3P_1$ intercombination transition of ${}^{88}$Sr near 689~nm, see Fig.~\ref{fig:experiment}(a). Using a magic-wavelength optical trap eliminates the spatially inhomogeneous frequency shift induced by the optical field, enabling our measurement of the CP shift. Furthermore, a magic-wavelength trap allows efficient transfer of atoms from a magneto-optical trap operating on the narrow-linewidth transition into the highly focused trap beam~\cite{Ido2000}. The optical lattice is formed by the interference of the incident laser and its retro-reflection from the dielectric surface, see Fig.~\ref{fig:experiment}(b). The dielectric surface used in this experiment is diced from a UV antireflection-coated fused-silica window (1~mm thick), whose reflectance at 914~nm was measured to be $0.188(11)$ at normal incidence. The phase shift upon reflection determines the distance of the closest lattice trapping site from the surface. From analyzing the optical thin-film coating layers, we calculate a reflection phase shift of $-2.62(3)$ rads (Appendix~\ref{app:Coating}). This causes the distance of the first lattice site to be at 189(2)~nm, which is smaller than a quarter of the lattice wavelength, around 228~nm. Finally, the total trapping potential show in Fig.~\ref{fig:experiment}(c) is then the sum of the CP potential and the optical lattice potential (Appendix~\ref{app:Lattice}).

We load the ultracold atoms into the lattice sites closest to the surface by modifying the bias magnetic field in the $-z$ direction, ${B}^{\text{bias}}_{\text{z}}$ , at the end of the magneto-optical trap cooling phase, while holding the atoms in the single-frequency red MOT [see Fig.~\ref{fig:timing}(a)]. Since the atoms are much closer to the surface, the change in the bias field magnitude is finer and happens over a $5$~ms period. Now, as the atoms are moving toward the surface, we abruptly turn on the lattice beam at a delay time $t_{\text{delay}}$ after the final magnetic field ramp end. The $t_{\text{delay}}$ determines the moving distance per unit of bias magnetic field change (Appendix \ref{app:Ramp}). 
If the lattice is turned on sooner, the atoms move up less. However, there are more atoms transferred into the lattice. The opposite is true if the lattice beam is turned on later. Here, the calibration for the distance per unit of bias magnetic field ramp is $860(25)\ \mathrm{\mu m}/\mathrm{G}$ for $t_{\text{delay}} = 5$~ms. The final value of ${B}^{\text{bias}}_{\text{z}}$ depends on the desired loading vertical locations of the lattice.

Finally, the lattice beam is overlapped with the single-frequency red MOT for $30$~ms. The atoms are then trapped in the optical lattice for $500$~ms before the probing sequence starts. After the red MOT beams are turned off, the ${B}^{\text{bias}}_{\text{z}}$ is set to zero, while we switch on the $y$-axis bias magnetic field, ${B}^{\text{bias}}_{\text{y}}$, for the spectroscopic measurement.

\section*{Narrow-linewidth spectroscopy}

We address technical requirement (ii) by using the 7.5-kHz-wide intercombination transition of $^{88}$Sr [Fig.~\ref{fig:experiment}(a)]. At the location of the first lattice site, the calculated CP shift on the 689-nm transition is $-15.6$~kHz (Appendix \ref{app:Theory}). The second trapping site is at half of the lattice wavelength away from the first site, or 645~nm from the surface. Because of the power-law decay of the CP potential as a function of distance, atoms at this location have their 689-nm transition shifted only by about $-283$~Hz, where it is not resolvable given the measured transition spectral linewidth of roughly 15~kHz. The 689-nm transition spectra of atoms at other lattice sites are indistinguishable. Thus, in our experiment, only the CP shift by the atoms trapped in the first lattice site is detectable. Furthermore, since the expected 689-nm transition frequency shift from the first lattice site is less than the spectral linewidth of the main peak, it is crucial to load many atoms into the first site while minimizing the number of atoms loaded into other trapping sites. Finally, we also set the optical lattice waist and power to operate in the Lamb-Dicke regime, see Appendix~\ref{app:Lattice}.

\section*{Fluorescence spectroscopy}

\begin{figure}[t!]
     \includegraphics[width=\columnwidth]{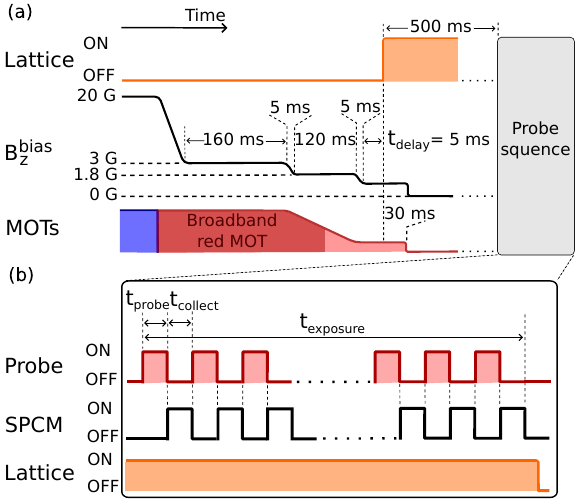}
     \caption{\textbf{Experimental timing sequence.} (a) The timing sequence shown starts during the red MOT stage with a multi-step process that brings the atoms to the test surface. An atom ramp-up sequence is mostly performed by stepping down the bias magnetic field in the $-z$ direction. Finally, loading the atoms in the red MOT into the optical lattice sites near the surface is carefully timed by experimenting with the $t_{\text{delay}}$ as explained in the main text. We choose $t_{\text{delay}} = 5$~ms, maximizing the number of atoms in the lattice trap. (b) Timing sequence for the time-gated fluorescence spectroscopy. The 689-nm probe beam is pulsed ON and OFF, while the single-photon counter module (SPCM) is enabled during the OFF period of the probe beam. This probing sequence begins $500$~ms after loading the atoms into the lattice, at the last stage of the timing sequence in (a). }
   \label{fig:timing}
\end{figure}

The final technical requirement (iii) required developing a novel imaging technique due to the physical constraints of our experimental setup. We perform time-gated fluorescence spectroscopy on the lattice-trapped atoms at the 689-nm transition. The 689-nm probe beam is linearly polarized along the $y$-axis, parallel to the quantization axis, which drives the $\pi$-transition. Its intensity is $I \approx I_{\text{sat}}$, where $I_{\text{sat}} = 3\ \mathrm{\mu W/cm^2}$ is the saturation intensity of the ${}^1S_0-{}^3P_1$ transition. To achieve Doppler-free and recoil-free spectra, the probe beam must be aligned along the axial direction of the optical lattice [Fig.~\ref{fig:experiment}(b)]. The probe beam propagates downward through the CP substrate and illuminates the atoms. It shines directly into the objective, which also collects the fluorescence photons from the atoms. After being collected by the objective, the fluorescence signal from the atoms is focused by a tube lens into a single-photon counter module (SPCM). It is worth noting that the bottom red MOT beam is combined with the fluorescence imaging path via a 50:50 beam splitter.

The time-gated probing sequence consists of a series of probing light pulses separated by a fluorescence photon-collection time [Fig. \ref{fig:timing}(b)]. Each probe pulse, which has the duration $t_{\text{probe}}$, excites a fraction of the atoms to the exited state ${}^3P_1\ (m_J=0)$. Then immediately after the probe pulse switch-off, the SPCM is enabled for $t_{\text{collect}}$ to capture the fluorescence photons while the atoms spontaneously decay back to the ground state. This excitation-detection cycle is repeated and lasts for $t_{\text{exposure}} = 80$~ms, where $t_{\text{probe}} = t_{\text{collect}} = 150\ \mathrm{\mu s}$ . The $t_{\text{probe}}$ duration is chosen to be long enough to avoid artificial broadening of the probe laser linewidth, while being short enough to prevent excessive heating of the atoms. On the other hand, $t_{\text{collect}}$ is ideally a few times longer than the ${}^3P_1$ excited state lifetime $\tau = 21\ \mathrm{\mu s}$. To avoid saturating the SPCM sensor, the SPCM is disabled (gate OFF) when the probe laser is pulsed on. In addition, we keep the SPCM disabled during MOT cooling phases and use a mechanical laser shutter to block all lights until the probing sequence begins. Narrow bandpass optical filters (CW at 690~nm) are used to block the intense scattered light from the lattice beam.  

To completely eliminate the magnetic field noise contribution in the spectroscopic measurement, we choose to probe the CP effect on the $\pi$-transition, ${}^1S_0-{}^3P_1\ (m_J=0)$. When performing the spectroscopy measurement, we set a bias magnetic field, ${B}^{\text{bias}}_{\text{y}}$, to be orthogonal to the lattice beam's polarization direction, $\mathbf{E}_{\text{trap}}$. In the experiment, the bias magnetic field points in the $+y$~direction, and the lattice beam is linearly polarized along the $x$-axis, where we define the $+z$~direction as the upward direction [Fig. \ref{fig:experiment}(a)]. The bias magnetic field direction defines the quantization axis, with a magnitude of ${B}^{\text{bias}}_{\text{y}} = 2.5$~G.

\section*{Results}

In our experiment, because optical access for atomic absorption imaging near the test surface is limited, determining the final magnitude of ${B}^{\text{bias}}_{\text{z}}$ is challenging. Fortunately, in an optical lattice, we expect only atoms at the first lattice site to give a resolvable spectroscopic signal. In practice, we progressively reduce the final ${B}^{\text{bias}}_{\text{z}}$ magnitude to find the CP shift signal from atoms in the first site. At ${B}^{\text{bias}}_{\text{z}} = 1.45$~G, our ``close'' (to the surface) spectroscopy scan reveals a secondary spectroscopy peak, which matches the expected CP shift. Whereas at ${B}^{\text{bias}}_{\text{z}} = 1.5$~G, or ``far,'' at about $43\ \mathrm{\mu m}$ lower by our distance calibration (Appendix~\ref{app:Ramp}) the spectrum only shows one peak. The optical lattice beam properties at these two locations were further charaterized and confirmed through sideband spectroscopy (Appendix~\ref{app:LatticeChar}).

\begin{figure}[t!]
     \includegraphics[width=\columnwidth]{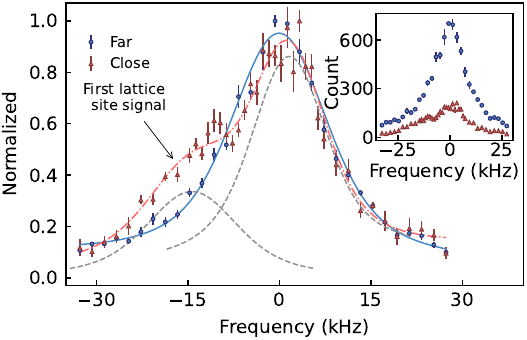}
     \caption{\textbf{Spectroscopic measurement of the Casimir-Polder force in the intermediate regime.} Normalized photon count data of the fluorescent spectroscopic scans across the narrow-linewidth ${}^1S_0-{}^3P_1\ (m_J=0)$ transition at two different optical lattice loading locations. We perform spectroscopy scans with the atoms loaded into the optical lattice at the far (blue circles) and the close (red triangles) locations. 
     It is clear that at the close lattice location, the spectroscopy data show another smaller peak that is red-detuned from the main peak. This signal comes from the atoms in the first lattice site, which experience a stronger CP potential. The transition frequency shift is the result of the differential CP shift between the ground state ${}^1S_0$ and the excited state ${}^3P_1\ (m_J=0)$. We fit the data with a single Voigt profile and a double Voigt profile for the far (solid blue) and the close (dash-dot red) data, respectively. The dashed curves illustrate individual fitted spectral peaks that constitute the second spectroscopic scan fit (dash-dot red). \textbf{Inset}: Raw photon count data of the scans. The photon count in the far data is about three times higher than the close data. }
     \label{fig:CP_data}
\end{figure}

Data in Fig. \ref{fig:CP_data} shows the fluorescent spectra of lattice trapped atoms at the far and the close locations. Naturally, due to more atom loss in the ramp-up process, the photon counts of the close data are several times smaller than the far one. To compare the fluorescent spectra across the two locations, we normalize the photon counts in each spectroscopy scan to its maximum and minimum values. At the far location, the spectroscopy data show little deviation from an atomic resonance spectrum. Here, we fit the data with a single Voigt profile to determine the center resonance frequency [${}^1S_0-{}^3P_1\ (m_J=0)$]. At the close location, in addition to the principal spectral peak centered at the resonance frequency measured at the far location, the data reveals a small red-detuned peak. We fit this spectroscopy data using the double-Voigt profile model. From the fit, the frequency shift of the secondary peak from the principal peak is $-15.8^{+1.7}_{-1.1}$~kHz, which is in good agreement with the QED model prediction shift of $-15.6$~kHz.
For comparison, in the short-distance approximation, the shift is $+1.9$~kHz; while in the long-distance approximation, the shift is undefined, because the $^3P_1(m_J=0)$ excited state  
is not trapped in the magic-wavelength lattice (Appendix~\ref{app:Theory}). The experimental observation rules out both of these approximations.

Due to concerns about strontium adsorption on the dielectric surface, we begin our data acquisition by aligning the lattice beam to a fresh area of the surface. We perform the first spectroscopy scan at the far position to minimize surface contamination. Then, we perform a second spectroscopy scan at the close position to probe the CP potential shift on a clean area. We subsequently run additional spectroscopy scans to monitor the behavior of the peaks in response to the strontium coating of the surface. We show these data in Appendix~\ref{app:SrAdsorption}.  

In this experiment, the major systematic error in the measured CP shift is the ac Stark shift induced by the optical lattice beam. We calculate the lattice peak intensities at the far and close location to be $60.9\ \mathrm{kW/cm^2}$ and $75.8\ \mathrm{kW/cm^2}$ respectively. The estimated differential polarizability between the ${}^1S_0$ and the ${}^3P_1$ state, $\Delta\alpha$, is $19(9)\ \mathrm{Hz/kW\cdot cm^{-2}}$ at the trapping wavelength. This gives the ac Stark shift between the far and close location to be $280(130)\ \mathrm{Hz}$. We anticipate a negligible frequency shift from the atomic density \cite{Ido2005}. We estimate the peak density of the lattice to be on the order of $10^{8}\ \mathrm{cm}^{-3}$. 

\section*{Conclusions}

In summary, we report direct spectroscopic measurement of the Casimir-Polder shift on the ${}^{88}$Sr intercombination transition at the intermediate range regime of a dielectric surface. At the atom-surface distance of $189$ nm, after propagating the uncertainty, the Casimir-Polder shift is  $-15.8^{+1.7}_{-1.1}$~kHz, which is in agreement with the QED model prediction of $-15.6$~kHz. Our experiment presents an excellent method for measurement of the Casimir-Polder potential, which shows a new benchmark for high precision. In future experiments, using a transparent substrate, the optical lattice can be formed by a separate counterpropagating beam. By physically translating one of the lattice beam, one can select an arbitrary atom-surface distance. 
For a reflective surface, a specially designed optical thin-film coating can provide high reflectivity at several magic wavelengths such as 473~nm \cite{Kestler2022}, 515~nm \cite{Cooper2018}, and 914~nm, while keeping good transmittance at 461~nm and 689~nm. With this setup, each magic wavelength gives the a different first lattice site location. Moreover, the reflection phase shift at each magic wavelength can be engineered by the coating to achieve greater distance tuning. In addition to the kHz-linewidth transition, the clock transition ${}^1 S_0-{}^3P_0$ with the sub-Hz natural linewidth is promising to improve the probing resolution of this method~\cite{Akatsuka2008}. The ability to accurately map the Casimir-Polder potential over a wider range in the intermediate regime is an essential tool for the development of novel quantum devices that trap atoms close to a surface.

\begin{acknowledgements}

We acknowledge support from the National Science Foundation Award No. 2412662.

\end{acknowledgements}

\bibliography{references}

@article{Casimir1948,
  title = {The Influence of Retardation on the {L}ondon-van der {W}aals Forces},
  author = {Casimir, H. B. G. and Polder, D.},
  journal = {Phys. Rev.},
  volume = {73},
  issue = {4},
  pages = {360--372},
  numpages = {0},
  year = {1948},
  month = {Feb},
  publisher = {American Physical Society},
  doi = {10.1103/PhysRev.73.360},
  url = {https://link.aps.org/doi/10.1103/PhysRev.73.360}
}

@article{Raskin1969,
  title = {Interaction between a Neutral Atomic or Molecular Beam and a Conducting Surface},
  author = {Raskin, D. and Kusch, P.},
  journal = {Phys. Rev.},
  volume = {179},
  issue = {3},
  pages = {712--721},
  numpages = {0},
  year = {1969},
  month = {Mar},
  publisher = {American Physical Society},
  doi = {10.1103/PhysRev.179.712}
}

@article{Shih1975,
  title = {Van der {W}aals forces between heavy alkali atoms and gold surfaces: Comparison of measured and predicted values},
  author = {Shih, A. and Parsegian, V. A.},
  journal = {Phys. Rev. A},
  volume = {12},
  issue = {3},
  pages = {835--841},
  numpages = {0},
  year = {1975},
  month = {Sep},
  publisher = {American Physical Society},
  doi = {10.1103/PhysRevA.12.835},
  url = {https://link.aps.org/doi/10.1103/PhysRevA.12.835}
}

@article{Anderson1988,
  title = {Measuring the van der {W}aals forces between a Rydberg atom and a metallic surface},
  author = {Anderson, A. and Haroche, S. and Hinds, E. A. and Jhe, W. and Meschede, D.},
  journal = {Phys. Rev. A},
  volume = {37},
  issue = {9},
  pages = {3594--3597},
  numpages = {0},
  year = {1988},
  month = {May},
  publisher = {American Physical Society},
  doi = {10.1103/PhysRevA.37.3594},
  url = {https://link.aps.org/doi/10.1103/PhysRevA.37.3594}
}

@article{Sandoghdar1992,
  title = {Direct measurement of the van der {W}aals interaction between an atom and its images in a micron-sized cavity},
  author = {Sandoghdar, V. and Sukenik, C. I. and Hinds, E. A. and Haroche, Serge},
  journal = {Phys. Rev. Lett.},
  volume = {68},
  issue = {23},
  pages = {3432--3435},
  numpages = {0},
  year = {1992},
  month = {Jun},
  publisher = {American Physical Society},
  doi = {10.1103/PhysRevLett.68.3432},
  url = {https://link.aps.org/doi/10.1103/PhysRevLett.68.3432}
}

@article{Landragin1996,
  title = {Measurement of the van der {W}aals Force in an Atomic Mirror},
  author = {Landragin, A. and Courtois, J.-Y. and Labeyrie, G. and Vansteenkiste, N. and Westbrook, C. I. and Aspect, A.},
  journal = {Phys. Rev. Lett.},
  volume = {77},
  issue = {8},
  pages = {1464--1467},
  numpages = {0},
  year = {1996},
  month = {Aug},
  publisher = {American Physical Society},
  doi = {10.1103/PhysRevLett.77.1464},
  url = {https://link.aps.org/doi/10.1103/PhysRevLett.77.1464}
}

@article{Shimizu2001,
  title = {Specular Reflection of Very Slow Metastable Neon Atoms from a Solid Surface},
  author = {Shimizu, Fujio},
  journal = {Phys. Rev. Lett.},
  volume = {86},
  issue = {6},
  pages = {987--990},
  numpages = {0},
  year = {2001},
  month = {Feb},
  publisher = {American Physical Society},
  doi = {10.1103/PhysRevLett.86.987},
  url = {https://link.aps.org/doi/10.1103/PhysRevLett.86.987}
}

@article{Pasquini2004,
  title = {Quantum Reflection from a Solid Surface at Normal Incidence},
  author = {Pasquini, T. A. and Shin, Y. and Sanner, C. and Saba, M. and Schirotzek, A. and Pritchard, D. E. and Ketterle, W.},
  journal = {Phys. Rev. Lett.},
  volume = {93},
  issue = {22},
  pages = {223201},
  numpages = {4},
  year = {2004},
  month = {Nov},
  publisher = {American Physical Society},
  doi = {10.1103/PhysRevLett.93.223201},
  url = {https://link.aps.org/doi/10.1103/PhysRevLett.93.223201}
}

@article{Marani2000,
  title = {Using atomic interference to probe atom-surface interactions},
  author = {Marani, Roberta and Cognet, Laurent and Savalli, Veronique and Westbrook, Nathalie and Westbrook, Christoph I. and Aspect, Alain},
  journal = {Phys. Rev. A},
  volume = {61},
  issue = {5},
  pages = {053402},
  numpages = {7},
  year = {2000},
  month = {Apr},
  publisher = {American Physical Society},
  doi = {10.1103/PhysRevA.61.053402},
  url = {https://link.aps.org/doi/10.1103/PhysRevA.61.053402}
}

@article{Harber2005,
  title = {Measurement of the {C}asimir-{P}older force through center-of-mass oscillations of a {B}ose-{E}instein condensate},
  author = {Harber, D. M. and Obrecht, J. M. and McGuirk, J. M. and Cornell, E. A.},
  journal = {Phys. Rev. A},
  volume = {72},
  issue = {3},
  pages = {033610},
  numpages = {6},
  year = {2005},
  month = {Sep},
  publisher = {American Physical Society},
  doi = {10.1103/PhysRevA.72.033610},
  url = {https://link.aps.org/doi/10.1103/PhysRevA.72.033610}
}

@article{Sukenik1993,
  title = {Measurement of the {C}asimir-{P}older force},
  author = {Sukenik, C. I. and Boshier, M. G. and Cho, D. and Sandoghdar, V. and Hinds, E. A.},
  journal = {Phys. Rev. Lett.},
  volume = {70},
  issue = {5},
  pages = {560--563},
  numpages = {0},
  year = {1993},
  month = {Feb},
  publisher = {American Physical Society},
  doi = {10.1103/PhysRevLett.70.560},
  url = {https://link.aps.org/doi/10.1103/PhysRevLett.70.560}
}

@article{Bender2010,
  title = {Direct Measurement of Intermediate-Range {C}asimir-{P}older Potentials},
  author = {Bender, H. and Courteille, Ph. W. and Marzok, C. and Zimmermann, C. and Slama, S.},
  journal = {Phys. Rev. Lett.},
  volume = {104},
  issue = {8},
  pages = {083201},
  numpages = {4},
  year = {2010},
  month = {Feb},
  publisher = {American Physical Society},
  doi = {10.1103/PhysRevLett.104.083201},
  url = {https://link.aps.org/doi/10.1103/PhysRevLett.104.083201}
}

@article{Garcion2021,
  title = {Intermediate-Range {C}asimir-{P}older Interaction Probed by High-Order Slow Atom Diffraction},
  author = {Garcion, C. and Fabre, N. and Bricha, H. and Perales, F. and Scheel, S. and Ducloy, M. and Dutier, G.},
  journal = {Phys. Rev. Lett.},
  volume = {127},
  issue = {17},
  pages = {170402},
  numpages = {6},
  year = {2021},
  month = {Oct},
  publisher = {American Physical Society},
  doi = {10.1103/PhysRevLett.127.170402},
  url = {https://link.aps.org/doi/10.1103/PhysRevLett.127.170402}
}

@article{Vetsch2010,
  title = {Optical Interface Created by Laser-Cooled Atoms Trapped in the Evanescent Field Surrounding an Optical Nanofiber},
  author = {Vetsch, E. and Reitz, D. and Sagu\'e, G. and Schmidt, R. and Dawkins, S. T. and Rauschenbeutel, A.},
  journal = {Phys. Rev. Lett.},
  volume = {104},
  issue = {20},
  pages = {203603},
  numpages = {4},
  year = {2010},
  month = {May},
  publisher = {American Physical Society},
  doi = {10.1103/PhysRevLett.104.203603},
  url = {https://link.aps.org/doi/10.1103/PhysRevLett.104.203603}
}

@article{Goban2012,
  title = {Demonstration of a State-Insensitive, Compensated Nanofiber Trap},
  author = {Goban, A. and Choi, K. S. and Alton, D. J. and Ding, D. and Lacro\^ute, C. and Pototschnig, M. and Thiele, T. and Stern, N. P. and Kimble, H. J.},
  journal = {Phys. Rev. Lett.},
  volume = {109},
  issue = {3},
  pages = {033603},
  numpages = {5},
  year = {2012},
  month = {Jul},
  publisher = {American Physical Society},
  doi = {10.1103/PhysRevLett.109.033603},
  url = {https://link.aps.org/doi/10.1103/PhysRevLett.109.033603}
}

@article{Goban2015,
  title = {Superradiance for Atoms Trapped along a Photonic Crystal Waveguide},
  author = {Goban, A. and Hung, C.-L. and Hood, J. D. and Yu, S.-P. and Muniz, J. A. and Painter, O. and Kimble, H. J.},
  journal = {Phys. Rev. Lett.},
  volume = {115},
  issue = {6},
  pages = {063601},
  numpages = {5},
  year = {2015},
  month = {Aug},
  publisher = {American Physical Society},
  doi = {10.1103/PhysRevLett.115.063601},
  url = {https://link.aps.org/doi/10.1103/PhysRevLett.115.063601}
}

@Article{Gonzalez-Tudela2015,
author={Gonz{\'a}lez-Tudela, A.
and Hung, C.-L.
and Chang, D. E.
and Cirac, J. I.
and Kimble, H. J.},
title={Subwavelength vacuum lattices and atom--atom interactions in two-dimensional photonic crystals},
journal={Nature Photon.},
year={2015},
month={May},
day={01},
volume={9},
number={5},
pages={320-325},
doi={10.1038/nphoton.2015.54},
url={https://doi.org/10.1038/nphoton.2015.54}
}

@article{Patterson2018,
  title = {Spectral asymmetry of atoms in the van der {W}aals potential of an optical nanofiber},
  author = {Patterson, B. D. and Solano, P. and Julienne, P. S. and Orozco, L. A. and Rolston, S. L.},
  journal = {Phys. Rev. A},
  volume = {97},
  issue = {3},
  pages = {032509},
  numpages = {8},
  year = {2018},
  month = {Mar},
  publisher = {American Physical Society},
  doi = {10.1103/PhysRevA.97.032509},
  url = {https://link.aps.org/doi/10.1103/PhysRevA.97.032509}
}

@article{Kestler2023,
  title = {State-Insensitive Trapping of Alkaline-Earth Atoms in a Nanofiber-Based Optical Dipole Trap},
  author = {Kestler, G. and Ton, K. and Filin, D. and Cheung, C. and Schneeweiss, P. and Hoinkes, T. and Volz, J. and Safronova, M.S. and Rauschenbeutel, A. and Barreiro, J.T.},
  journal = {PRX Quantum},
  volume = {4},
  issue = {4},
  pages = {040308},
  numpages = {19},
  year = {2023},
  month = {Oct},
  publisher = {American Physical Society},
  doi = {10.1103/PRXQuantum.4.040308},
  url = {https://link.aps.org/doi/10.1103/PRXQuantum.4.040308}
}

@article{Zhou2024,
  title = {Trapped Atoms and Superradiance on an Integrated Nanophotonic Microring Circuit},
  author = {Zhou, Xinchao and Tamura, Hikaru and Chang, Tzu-Han and Hung, Chen-Lung},
  journal = {Phys. Rev. X},
  volume = {14},
  issue = {3},
  pages = {031004},
  numpages = {11},
  year = {2024},
  month = {Jul},
  publisher = {American Physical Society},
  doi = {10.1103/PhysRevX.14.031004},
  url = {https://link.aps.org/doi/10.1103/PhysRevX.14.031004}
}

@Article{Chang2014,
author={Chang, D. E.
and Sinha, K.
and Taylor, J. M.
and Kimble, H. J.},
title={Trapping atoms using nanoscale quantum vacuum forces},
journal={Nat. Commun.},
year={2014},
month={Jul},
day={10},
volume={5},
number={1},
pages={4343},
doi={10.1038/ncomms5343},
url={https://doi.org/10.1038/ncomms5343}
}

@article{Le2007,
  title = {Spontaneous radiative decay of translational levels of an atom near a dielectric surface},
  author = {Le Kien, Fam and Hakuta, K.},
  journal = {Phys. Rev. A},
  volume = {75},
  issue = {1},
  pages = {013423},
  numpages = {13},
  year = {2007},
  month = {Jan},
  publisher = {American Physical Society},
  doi = {10.1103/PhysRevA.75.013423},
  url = {https://link.aps.org/doi/10.1103/PhysRevA.75.013423}
}

@article{Le2022,
  title = {Repulsive {C}asimir-{P}older potentials of low-lying excited states of a multilevel alkali-metal atom near an optical nanofiber},
  author = {Le Kien, Fam and Kornovan, D. F. and Nic Chormaic, S\'{\i}le and Busch, Thomas},
  journal = {Phys. Rev. A},
  volume = {105},
  issue = {4},
  pages = {042817},
  numpages = {14},
  year = {2022},
  month = {Apr},
  publisher = {American Physical Society},
  doi = {10.1103/PhysRevA.105.042817},
  url = {https://link.aps.org/doi/10.1103/PhysRevA.105.042817}
}

@article{Stern_2011, 
doi = {10.1088/1367-2630/13/8/085004}, 
url = {https://doi.org/10.1088/1367-2630/13/8/085004}, 
year = {2011}, 
month = {aug}, 
publisher = {}, 
volume = {13}, 
number = {8}, 
pages = {085004}, 
author = {Stern, N P and Alton, D J and Kimble, H J}, 
title = {Simulations of atomic trajectories near a dielectric surface}, 
journal = {New Journal of Physics}
}

@article{Hung_2013,
doi = {10.1088/1367-2630/15/8/083026},
url = {https://doi.org/10.1088/1367-2630/15/8/083026},
year = {2013},
month = {aug},
publisher = {IOP Publishing},
volume = {15},
number = {8},
pages = {083026},
author = {Hung, C-L and Meenehan, S M and Chang, D E and Painter, O and Kimble, H J},
title = {Trapped atoms in one-dimensional photonic crystals},
journal = {New Journal of Physics}
}

@article{Kestler2022,
  title = {Magic wavelengths of the Sr ($5{s}^{2} {}^{1}S_{0}$--$5s5p {}^{3}P_{1}$) intercombination transition near the $5s5p {}^{3}P_{1}$--$5{p}^{2} {}^{3}P_{2}$ transition},
  author = {Kestler, Grady and Ton, Khang and Filin, Dmytro and Safronova, Marianna S. and Barreiro, Julio T.},
  journal = {Phys. Rev. A},
  volume = {105},
  issue = {1},
  pages = {012821},
  numpages = {7},
  year = {2022},
  month = {Jan},
  publisher = {American Physical Society},
  doi = {10.1103/PhysRevA.105.012821},
  url = {https://link.aps.org/doi/10.1103/PhysRevA.105.012821}
}

@article{Cooper2018,
  title = {Alkaline-Earth Atoms in Optical Tweezers},
  author = {Cooper, Alexandre and Covey, Jacob P. and Madjarov, Ivaylo S. and Porsev, Sergey G. and Safronova, Marianna S. and Endres, Manuel},
  journal = {Phys. Rev. X},
  volume = {8},
  issue = {4},
  pages = {041055},
  numpages = {19},
  year = {2018},
  month = {Dec},
  publisher = {American Physical Society},
  doi = {10.1103/PhysRevX.8.041055},
  url = {https://link.aps.org/doi/10.1103/PhysRevX.8.041055}
}

@article{Ido2003,
  title = {Recoil-Free Spectroscopy of Neutral Sr Atoms in the {L}amb-{D}icke Regime},
  author = {Ido, Tetsuya and Katori, Hidetoshi},
  journal = {Phys. Rev. Lett.},
  volume = {91},
  issue = {5},
  pages = {053001},
  numpages = {4},
  year = {2003},
  month = {Jul},
  publisher = {American Physical Society},
  doi = {10.1103/PhysRevLett.91.053001},
  url = {https://link.aps.org/doi/10.1103/PhysRevLett.91.053001}
}

@article{Ido2005,
  title = {Precision Spectroscopy and Density-Dependent Frequency Shifts in Ultracold Sr},
  author = {Ido, Tetsuya and Loftus, Thomas H. and Boyd, Martin M. and Ludlow, Andrew D. and Holman, Kevin W. and Ye, Jun},
  journal = {Phys. Rev. Lett.},
  volume = {94},
  issue = {15},
  pages = {153001},
  numpages = {4},
  year = {2005},
  month = {Apr},
  publisher = {American Physical Society},
  doi = {10.1103/PhysRevLett.94.153001},
  url = {https://link.aps.org/doi/10.1103/PhysRevLett.94.153001}
}

@article{Leibfried2003,
  title = {Quantum dynamics of single trapped ions},
  author = {Leibfried, D. and Blatt, R. and Monroe, C. and Wineland, D.},
  journal = {Rev. Mod. Phys.},
  volume = {75},
  issue = {1},
  pages = {281--324},
  numpages = {0},
  year = {2003},
  month = {Mar},
  publisher = {American Physical Society},
  doi = {10.1103/RevModPhys.75.281},
  url = {https://link.aps.org/doi/10.1103/RevModPhys.75.281}
}

@article{Ido2000,
  title = {Optical-dipole trapping of Sr atoms at a high phase-space density},
  author = {Ido, Tetsuya and Isoya, Yoshitomo and Katori, Hidetoshi},
  journal = {Phys. Rev. A},
  volume = {61},
  issue = {6},
  pages = {061403},
  numpages = {4},
  year = {2000},
  month = {May},
  publisher = {American Physical Society},
  doi = {10.1103/PhysRevA.61.061403},
  url = {https://link.aps.org/doi/10.1103/PhysRevA.61.061403}
}

@article{Snigirev2019,
  title = {Fast and dense magneto-optical traps for strontium},
  author = {Snigirev, S. and Park, A. J. and Heinz, A. and Bloch, I. and Blatt, S.},
  journal = {Phys. Rev. A},
  volume = {99},
  issue = {6},
  pages = {063421},
  numpages = {9},
  year = {2019},
  month = {Jun},
  publisher = {American Physical Society},
  doi = {10.1103/PhysRevA.99.063421},
  url = {https://link.aps.org/doi/10.1103/PhysRevA.99.063421}
}

@article{Loftus2004,
  title = {Narrow line cooling and momentum-space crystals},
  author = {Loftus, Thomas H. and Ido, Tetsuya and Boyd, Martin M. and Ludlow, Andrew D. and Ye, Jun},
  journal = {Phys. Rev. A},
  volume = {70},
  issue = {6},
  pages = {063413},
  numpages = {14},
  year = {2004},
  month = {Dec},
  publisher = {American Physical Society},
  doi = {10.1103/PhysRevA.70.063413},
  url = {https://link.aps.org/doi/10.1103/PhysRevA.70.063413}
}

@article{FRANTAMgF2,
title = {Universal dispersion model for characterization of optical thin films over wide spectral range: Application to magnesium fluoride},
journal = {Applied Surface Science},
volume = {421},
pages = {424-429},
year = {2017},
issn = {0169-4332},
doi = {https://doi.org/10.1016/j.apsusc.2016.09.149},
url = {https://www.sciencedirect.com/science/article/pii/S016943321632030X},
author = {Daniel Franta and David Nečas and Angelo Giglia and Pavel Franta and Ivan Ohlídal}
}

@article{FrantaHfO2,
author = {Daniel Franta and David Nečas and Ivan Ohlídal},
journal = {Appl. Opt.},
number = {31},
pages = {9108--9119},
publisher = {Optica Publishing Group},
title = {Universal dispersion model for characterization of optical thin films over a wide spectral range: application to hafnia},
volume = {54},
month = {Nov},
year = {2015},
url = {https://opg.optica.org/ao/abstract.cfm?URI=ao-54-31-9108},
doi = {10.1364/AO.54.009108},
}

@article{FrantaSiO2,
author = {D. Franta and D. Nečas and I. Ohlídal and A. Giglia},
title = {Optical characterization of {S}i{O}2 thin films using universal dispersion model over wide spectral range},
journal = {Proc. SPIE},
volume = {9890}, 
pages = {989014},
year = {2016},
doi = {10.1117/12.2227580},
url = {https://www.spiedigitallibrary.org/conference-proceedings-of-spie/9890/1/Optical-characterization-of-SiO2-thin-films-using-universal-dispersion-model/10.1117/12.2227580.short}
}

@article{Mitroy10,
  title = {Theory and applications of atomic and ionic polarizabilities},
  author = {J. Mitroy and M. S. Safronova and Charles W Clark},
  journal = {{J.\ Phys.\ B: At.\ Mol.\ Opt.\ Phys.}},
  volume = {43},
  pages = {202001},
  year = {2010},
  doi = {10.1088/0953-4075/43/20/202001}
}

@article{Safronova13,
  title = {Blackbody-radiation shift in the Sr optical atomic clock},
  author = {M. S. Safronova and S. G. Porsev and U. I. Safronova and M. G. Kozlov and Charles W. Clark},
  journal = {Phys.\ Rev.\ A},
  volume = {87},
  pages = {012509},
  year = {2013},
  doi = {10.1103/PhysRevA.87.012509}
}

@book{Pedrotti,
  title = {Introduction to Optics},
  author = {Frank L. Pedrotti and Leno M. Pedrotti and Leno S. Pedrotti},
  edition = {3rd},
  publisher = {Cambridge University Press},
  year = {2017},
  city = {New York}
}

@article{Abeles50a,
  title = {Recherches sur la propagation des ondes électromagnétiques sinusoïdales dans les milieux stratifiés. Application aux couches minces},
  author = {Florin Abelès},
  journal = {Ann. Phys. (Paris)},
  volume = {12},
  pages = {596},
  year = {1950},
  doi = {10.1051/anphys/195012050596}
}

@article{Abeles50b,
  title = {Recherches sur la propagation des ondes électromagnétiques sinusoïdales dans les milieux stratifiés. Application aux couches minces. Deuxiéme partie},
  author = {Florin Abelès},
  journal = {Ann. Phys. (Paris)},
  volume = {12},
  pages = {706},
  year = {1950},
  doi = {10.1051/anphys/195012050706}
}

@article{Abeles50c,
  title = {La théorie générale des couches minces},
  author = {Florin Abelès},
  journal = {{J. Phys. Rad.}},
  volume = {11},
  pages = {307},
  year = {1950},
  doi = {10.1051/jphysrad:01950001107030700}
}

@misc{refractiveindex,
  howpublished = {\url{https://refractiveindex.info}}
}

@article{Bimonte10,
  title = {Generalized Kramers-Kronig transform for {C}asimir effect computations},
  author = {Giuseppe Bimonte},
  journal = {\pra},
  volume = {81},
  pages = {062501},
  year = {2010},
  doi = {10.1103/PhysRevA.81.062501}
}

@article{McLachlan63a,
  title = {Retarded dispersion forces between molecules},
  author = {A. D. McLachlan},
  journal = {{Proc. R. Soc. Lond., Ser. A, Math. Phys. Sci.}},
  volume = {271},
  pages = {387},
  year = {1963},
  doi = {10.1098/rspa.1963.0025}
}

@article{McLachlan63b,
  title = {Van der {W}aals forces between an atom and a surface},
  author = {A. D. McLachlan},
  journal = {{Mol.\ Phys.}},
  volume = {7},
  pages = {381},
  year = {1963},
  doi = {10.1080/00268976300101141}
}

@article{Wylie84,
  title = {Quantum electrodynamics near an interface,},
  author = {J. M. Wylie and J. E. Sipe},
  journal = {\pra},
  volume = {30},
  pages = {1185},
  year = {1984},
  doi = {10.1103/PhysRevA.30.1185}
}

@article{Wylie85,
  title = {Quantum electrodynamics near an interface. II},
  author = {J. M. Wylie and J. E. Sipe},
  journal = {\pra},
  volume = {32},
  pages = {2030},
  year = {1985},
  doi = {10.1103/PhysRevA.32.2030}
}

@book{Vogel,
  title = {Quantum Optics},
  author = {Werner Vogel and Dirk-Gunnar Welsch},
  edition = {3rd},
  publisher = {Wiley-VCH},
  year = {2006},
  city = {Weinheim}
}

@article{Malitson65,
  title = {Interspecimen Comparison of the Refractive Index of Fused Silica},
  author = {I. H. Malitson},
  journal = {{J. Opt.\ Soc.\ Am.}},
  volume = {55},
  pages = {1205},
  year = {1965},
  doi = {10.1364/JOSA.55.001205}
}

@incollection{Johnson_2011,
  author    = {Johnson, Steven G.},
  title     = {Numerical Methods for Computing {C}asimir Interactions},
  booktitle = {Casimir Physics},
  editor    = {Dalvit, Diego and Milonni, Peter and Roberts, David and da Rosa, Felipe},
  series    = {Lecture Notes in Physics},
  volume    = {834},
  pages     = {175},
  publisher = {Springer Berlin Heidelberg},
  address   = {Berlin, Heidelberg},
  year      = {2011},
  doi       = {10.1007/978-3-642-20288-9_7}
}

@article{Reid_2013,
  title = {Fluctuating surface currents: An algorithm for efficient prediction of {C}asimir interactions among arbitrary materials in arbitrary geometries},
  author = {Reid, M. T. Homer and White, Jacob and Johnson, Steven G.},
  journal = {Phys. Rev. A},
  volume = {88},
  pages = {022514},
  numpages = {24},
  year = {2013},
  month = {Aug},
  publisher = {American Physical Society},
  doi = {10.1103/PhysRevA.88.022514}
}

@article{Reid_2009,
  title = {Efficient Computation of {C}asimir Interactions between Arbitrary 3D Objects},
  author = {Reid, M. T. Homer and Rodriguez, Alejandro W. and White, Jacob and Johnson, Steven G.},
  journal = {Phys. Rev. Lett.},
  volume = {103},
  pages = {040401},
  numpages = {4},
  year = {2009},
  month = {Jul},
  publisher = {American Physical Society},
  doi = {10.1103/PhysRevLett.103.040401},
  url = {https://link.aps.org/doi/10.1103/PhysRevLett.103.040401}
}

@article{Akatsuka2008,
title    = "Optical lattice clocks with non-interacting bosons and fermions",
author   = "Akatsuka, Tomoya and Takamoto, Masao and Katori, Hidetoshi",
journal  = "Nature Physics",
volume   =  4,
number   =  12,
pages    = "954--959",
month    =  dec,
year     =  2008,
doi      = {10.1038/nphys1108}
}

\begin{appendix}
\section{Theoretical Casimir-Polder potential calculation} \label{app:Theory}
\edef\ket#1{|{#1}\rangle}
\edef\bra#1{\langle{#1}|}
\def\VCP{V_\mathrm{\scriptscriptstyle CP}}
\def\eqnarr#1#2{  
\renewcommand{\arraystretch}{#1}
  \setlength\arraycolsep{0ex}
  \begin{array}{rcl}
    #2
  \end{array}
}
\def\ds{\displaystyle}
\def\arreq{&{}={}&\ds }
\def\arrnone{&&\ds }
\def\intzinf{\int_{0}^\infty\!\!}
\def\kT{k_{\scriptscriptstyle \mathsf{T}}}
\def\shrinkage{1mu}
\def\vecsign{\mathchar"017E}
\def\dvecsign{\smash{\stackon[-1.99pt]{\mkern-\shrinkage\vecsign}{\rotatebox{180}{$\mkern-\shrinkage\vecsign$}}}}

The Casimir-Polder (CP) force is calculated from 
\begin{equation}
  \VCP = \VCP^{(1)}+\VCP^{(2)}.
\end{equation}
The main contribution is \cite{McLachlan63a, McLachlan63b, Wylie84, Wylie85, Vogel}
\begin{widetext}
\begin{equation}
  \eqnarr{2.1}{
  \VCP^{(1)}(z)
  \arreq\frac{\hbar}{16\pi^2\epsilon_0c^2}\intzinf \!ds\,s^2
    \bigg(\frac{\left[\alpha_{xx}(is)+\alpha_{yy}(is)\right]}{2}+\alpha_{zz}(is)\bigg)
    \intzinf\!d\kT\,
    \frac{\kT}{\kappa}
    \left[
    \bigg( r_\perp
    +\Big(1+\frac{2\kT^{\,2}c^2}{s^2}\Big)r_\parallel\bigg)\!
    \right]    \!  e^{-2\kappa z}
  \\\arrnone{}\hspace{5mm}
    +\frac{\hbar}{16\pi^2\epsilon_0c^2}\intzinf \!ds\,s^2
    \bigg(\frac{\left[\alpha_{xx}(is)+\alpha_{yy}(is)\right]}{2}-\alpha_{zz}(is)\bigg)
    \intzinf\!d\kT\,
    \frac{\kT}{\kappa}
    \big( r_\perp+r_\parallel
    \big)
         e^{-2\kappa z},
  }
  \label{asymmetricatomCPdielectric}
\end{equation}
\end{widetext}
where
\begin{equation}
 \kappa(s,\kT):=\sqrt{s^2/c^2+\kT^{\,2}},
 \label{kappadef}
\end{equation}
$z$ is the distance between the atom and surface, and
the imaginary-frequency atomic polarizability tensor for the level in question ($\ket{n}$) is
\begin{equation}
  \alpha_{\mu\nu}(is)
   =\sum_j\frac{2\omega_{jn}\bra{n}{d}_\mu\ket{j}
   \bra{j}{d}_\nu\ket{n}}
   {\hbar(\omega_{jn}^{\,2}+s^2)}.
   \label{KHalpha}
\end{equation}
The $r_\perp$ and $r_\parallel$ are the optical reflection coefficients for
transverse-electric (TE) and transverse-magnetic (TM) polarization, respectively,
defined such that they both reduce to $-1$ for normal incidence on a perfect conductor.
The first and second terms of Eq.~(\ref{asymmetricatomCPdielectric}) represent the 
contributions due to the symmetric and asymmetric parts of the polarizability, respectively.
The other contribution to the CP potential is necessary for (relative) excited states,
\begin{equation}
    \VCP^{(2)} (z)
    =
    -\sum_j\Theta(\omega_{nj})
     \bra{n}{d}_\mu\ket{j}
     \bra{j}{d}_\nu\ket{n}\,
     \mathrm{Re}[G_{\mu\nu}^{\mathrm{(s)}}(\mathbf{r},\mathbf{r},\omega_{jn})],
\end{equation}
where the nonzero components of scattering part $G_{\mu\nu}^{(s)}$
of the Green tensor are given for a planar reflector at distance
$z$ by
\begin{widetext}
\begin{equation}
  \eqnarr{2.2}{
  G_{zz}^{(s)}(z,z,\omega)\arreq
    -\frac{i}{4\pi\epsilon_0}
     \intzinf d\kT\,\frac{\kT^{\,3}}{k_z}
       r_\parallel(\theta,\omega)\,e^{i2k_zz}
       \\
  G_{xx}^{(s)}(z,z,\omega)=
  G_{yy}^{(s)}(z,z,\omega)\arreq
    \frac{i}{8\pi\epsilon_0}
     \intzinf d\kT\frac{\kT}{k_z}
     \left[k_z^{\,2}r_\parallel(\theta,\omega)
       +\frac{\omega^2}{c^2}r_\perp(\theta,\omega)\right]e^{i2k_zz},
  }
\end{equation}
\end{widetext}
and $\smash{k_z=\sqrt{\omega^2/c^2-\kT^{\,2}}}$.
This term applies only to the 689-nm transition in $^{88}$Sr, for the 
excited state in that transition.

Using the static polarizabilities from Ref.~\cite{Safronova13} for the ${}^1S_0$ (ground)
state, we find that the polarizability computed from the level sum is about $3.5\%$ too small.
Similarly, using the theoretical static polarizabilities from Ref.~\cite{Mitroy10} 
for the ${}^3P_1$ (excited) state, our polarizability is about 
$15.5\%$ too small.
These static polarizabilities apply to the far-field CP potential.
For these experiments, which lie in the intermediate regime, we do not apply any corrections
to the calculated potentials.

The CP potential depends on the complex index of refraction via the 
reflection coefficients for the two polarizations.
Their calculation is detailed more in Methods~\ref{app:Coating},
but because the coefficients depend on the imaginary frequency $is$,
they depend on the refractive index of the same argument:
\begin{equation}
     \tilde{n}(is)=
         \epsilon_0 +\frac{2}{\pi}\int_{0}^{\infty}
            \frac{\omega'\mathrm{Im}[\tilde{n}(\omega')]}{\omega'^2+s^2}\,
            d\omega'.
    \label{refractiveindex}
\end{equation}
This function is real-valued and monotonically decreasing.  It also depends only on the 
imaginary (absorptive) part of the refractive index along the real line.
It is difficult to quantify the error from truncating the measured frequency dependence
of the refractive index, but the data are available over a wide range
($0.115-125$~$\mu$m for HfO$_2$ \cite{FRANTAHfO2}, 
$0.0275-125$~$\mu$m for MgF$_2$ \cite{FRANTAMgF2},
and an analytical model for fused silica \cite{Malitson65};
the data were downloaded from \cite{refractiveindex}).
Further, on the high-frequency side, the convergence of 
Eq.~(\ref{refractiveindex}) is helped by the falloff of $\tilde{n}(\omega)$ and
the factor of $\omega'^2$ in the denominator;
and on the low-frequency side the data extend below the lowest resonance
frequencies of the materials. Ref.~\cite{Bimonte10} calls the errors in 
these cases ``negligible'' (certainly in comparison to the error induced by
the polarizabilities).

\begin{figure}[t!]
  \includegraphics[width=\columnwidth]{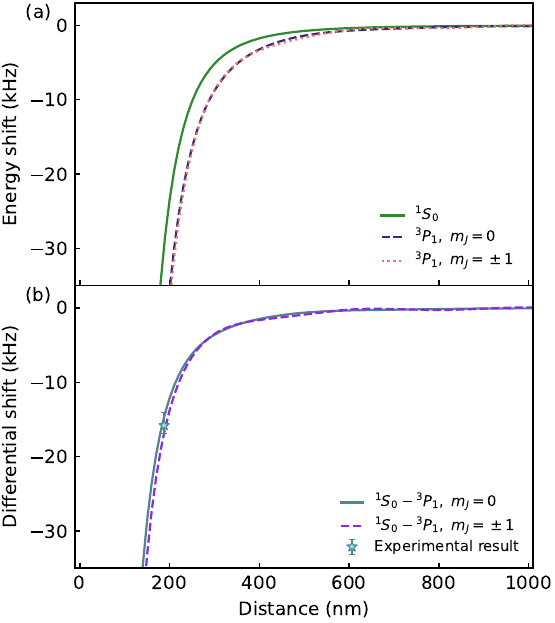}
  \caption{\textbf{Theoretical CP shift}
  (a) The calculated CP shifts of the ${}^1S_0$
  and ${}^3P_1$ levels of $^{88}$Sr.
  (b) The corresponding differential shifts.
  The experiment deals with the ${}^3P_1\; (m=0)$ state, but the ${}^3P_1\;(m=\pm 1)$ states are shown as well, to show that the difference in this setup is fairly small (below experimental resolution).
  }
   \label{fig:theorydata}
\end{figure}

The calculated absolute and differential level shifts 
due to the CP effect
are shown in Fig.~\ref{fig:theorydata}.
In order to convert this into an experimental prediction, we superpose
the CP potentials with the potential of the magic-wavelength
lattice (which depends on the phase shift as discussed in the next section). 
We then find the motional ground-state energies via the shooting method;
the difference between these energies yields the prediction for the experiment,
which is $-15.6$~kHz.

Also in the Results section of the main manuscript, for comparison, we report the CP
transition shifts in the short-range and long-range approximations.
In the short-range approximation, the CP potential
is dominated by short-wavelength evanescent modes.
Thus, the imaginary frequency $s$ is negligible compared to $\kT$,
Eq.~(\ref{kappadef}) is replaced by $\kappa\approx\kT$,
the reflection coefficient $r_\parallel$ dominates $r_\perp$,
and the second term of Eq.~(\ref{asymmetricatomCPdielectric}) is
negligible compared to the first.
The only dependence on the imaginary frequency is through the
frequency-dependent polarizabilities and the refractive
index, which we take to be of MgF$_2$ (being the outermost coating
layer; see Methods~\ref{app:Coating}). 
In this regime, the remaining component of the Casimir--Polder
potential scales as $z^{-3}$.
However, the calculation of the classical 
radiative component of the excited state
[$\smash{\VCP^{(2)} (z)}$] proceeds as in the general case, because there
is no sensible short-wavelength approximation.
The shift we obtain numerically in this approximation is
$+1.9$~kHz.

In the long-range approximation, the longest electromagnetic wavelengths
dominate the potential. Thus, we ignore the optical coating and treat
the interface as if it were only with the fused-silica substrate; 
for the dc limit of 
the refractive index, we use the value at a wavelength 
125~$\mu$m reported
in Ref.~\cite{FrantaSiO2} of about 2.038.  We also use the dc polarizabilities
for the strontium atom.  
In this regime, both terms of Eq.~(\ref{asymmetricatomCPdielectric})
scale as $z^{-4}$.
Again, the calculation of the classical radiative component proceeds
as in the general case, since there is no appropriate
long-wavelength approximation.
In this approximation, we do not obtain a shift due to the Casimir--Polder
effect, because the $^3P_1(m_J=0)$ excited state is not trapped in the magic-wavelength optical lattice (that is, there is no side peak as in Fig.~\ref{fig:CP_data}).

\section{Ultracold atoms preparation} \label{app:Prep}
We prepare an ultracold ${}^{88}$Sr atom cloud with the well-known method of using two-stage magneto-optical traps (MOTs) operating on the dominant 461-nm transition (blue MOT) and then on the narrow-linewidth 689-nm transition (red MOT) \cite{Loftus2004,Snigirev2019}. In the blue MOT cooling stage, the atom cloud forms approximately $3$~mm from the test surface. Here, we apply a $20$~G bias magnetic field in the $-z$ direction, ${B}^{\text{bias}}_{\text{z}}$, to shift the MOT quadrupole magnetic field center down. We keep the blue MOT far below to prevent strontium contamination over a large area of the test surface. Upon transferring the atoms into the narrow-linewidth cooling red MOT, the quadrupole magnetic field gradient is quickly ramped down from about $60$~G/cm  to about $12$~G/cm. In response to that change, the ${B}^{\text{bias}}_{\text{z}}$ magnitude is reduced to $3$~G. This magnetic field configuration is maintained for $160$~ms during the broadband red MOT phase. After the atoms have been cooled and accumulated into a more compact cloud, the bias field is again lowered to $1.8$~G over $5$~ms, further raising the MOT cloud. The timing diagram of the atom ramp-up sequence is depicted in Fig. \ref{fig:timing}(a). In the single-frequency red MOT phase, the red MOT beams' laser detuning is set to a relatively small value of $\delta = -190$~kHz. This detuning sets the red MOT cloud position at around $300\ \mathrm{\mu m}$ from the test surface. This single-frequency red MOT cloud has about $5\times10^4$ atoms, which is at a temperature of approximately $1\ \mathrm{\mu K}$. This atom cloud resembles a disk shape approximately $70\ \mathrm{\mu m}$ thick and $250\ \mathrm{\mu m}$ in diameter.

\section{Optical coating measurement} \label{app:Coating}

\begin{figure}[t!]
     \includegraphics[width=\columnwidth]{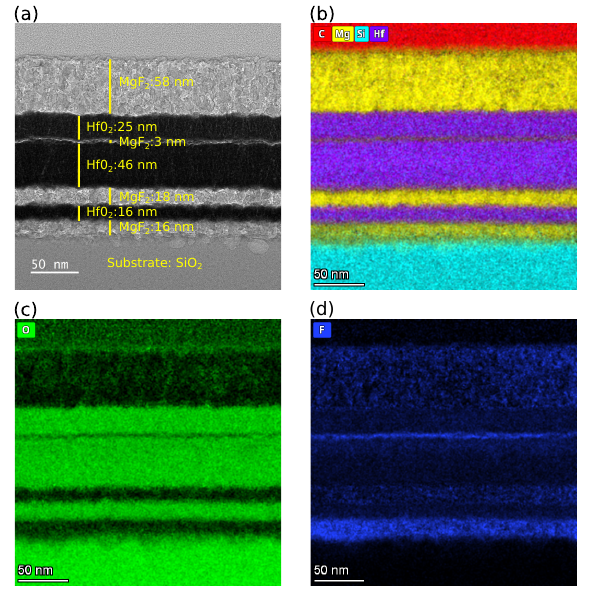}
     \caption{\textbf{TEM of CP test surface} (a) The TEM image shows the individual layer thickness of the optical thin-film coating of our CP test surface. The coating comprises seven alternating layers of $\mathrm{MgF_2}$ and $\mathrm{HfO_2}$ with varying thicknesses. (b), (c), and (d) show the EDS maps of the coating layers. These reveal the material composition. The top layer of carbon (red) in (b) was only introduced as a protective layer for the TEM imaging procedure. It is absent in the CP test surface.}
   \label{fig:temscan}
\end{figure}

The CP potential calculation requires precise knowledge of the surface dielectric coating. Here, the UV anti-reflection optical coating is analyzed under transmission electron microscope (TEM) with energy dispersive X-ray spectroscopy (EDS) by Covalent at Sunnyvale, CA, USA (Fig. \ref{fig:temscan}). Besides the calculation of the CP potential as a function of distance (Methods~\ref{app:Theory}), we calculate the distance of the first lattice site using the optical coating data. 

The distance of the first lattice site strongly depends on the reflection phase shift of the 914-nm beam. Due to the optical coating, the reflection phase shift is different from $\pi$. We analyze the propagation of the 914-nm beam through the thin-film stack layers by using the transfer matrix method of Abelès \cite{Abeles50a, Abeles50b, Abeles50c, Pedrotti}:

\begin{equation}
    M_s = M_7\cdot M_6\cdot ...\cdot M_1 = \begin{pmatrix} A & B \\ C & D \end{pmatrix}.
\end{equation}
Here, $M_1$ to $M_7$ are the transfer matrices corresponded to the seven layers. The transfer matrix takes the form of
\begin{equation}
    M_j =  \begin{pmatrix} \cos{\delta_j} & \frac{-i\sin{\delta_j}}{\alpha_j} \\ -i\alpha_i \sin{\delta_j}  & \cos{\delta_j}  \end{pmatrix},
\end{equation}
where $\delta_j = \frac{2\pi}{\lambda_0}\tilde{n}_j d_j \cos{\theta_{tj}}$ is the phase shift after transmitting through the thin-film layer $j$, the admittance $\alpha_j = \frac{\tilde{n}_j}{\eta_0}\cos{\theta_{tj}}$ for s-polarization (TE), and $\alpha_j = \frac{\tilde{n}_j}{\eta_0\cos{\theta_{tj}}}$ for p-polarization (TM). In this expression, $\tilde{n} =  n +  i\kappa$ is the complex refractive index. Lastly, $d_j$ is the thickness of the layer $j$.  
Upon getting the total transfer matrix, we have the complex reflectivity as 
\begin{equation}
    r = \frac{\alpha_i A + \alpha_i \alpha_s B - C - \alpha_s D}{\alpha_i A + \alpha_i \alpha_s B + C + \alpha_s D} .
\end{equation}
Here $\alpha_i$ and $\alpha_s$ are the admittances of the incident medium and the substrate, respectively. The reflection phase shift is then $\phi = \text{arctan}\frac{\text{Im}[r]}{\text{Re}[r]}$.

We calculate the reflection phase shift from the thin-film coating data. We simulate a 1-nm random thickness variation uniformly distributed across the layers to account for the coating fabrication tolerance. The resulting reflection phase shift is $-2.62(3)$ rads. In addition, we calculate the theoretical reflectance at 914 nm to be 0.178. Our reflectance measurement of 0.188(11) confirms this result.

\section{Optical lattice setup} \label{app:Lattice}

The lattice beam is launched from the bottom of the science chamber through a long working distance objective (Mitutoyo 20X N Plan Apo NIR, NA~=~0.4). We use a cavity-locked Ti:sapphire laser to generate a 914-nm optical lattice beam that propagates through free space to the objective. It typically has $430$~mW of power measured before the objective. After transmitting through the objective and the vacuum chamber's window, the power of the incident lattice beam reaching the cold strontium atom cloud is approximately $270$~mW. The beam is focused onto the CP test surface with a $1/e$ waist radius of $\omega_0 \approx 20\ \mathrm{\mu m}$. We perform resolved-sideband spectroscopy to better characterize the optical lattice (Methods \ref{app:Lattice}). From the heating sideband, we measure the axial trap frequency to be $\nu_z = 176(4)$~kHz, consistent with the simulated result. From here, we calculate the trap depth of the first lattice trapping site and the adjacent sites to be $37\ \mathrm{\mu K}$ and $148\ \mathrm{\mu K}$ respectively [Fig. \ref{fig:experiment}(c)]. The corresponding radial trapping frequency is $1900$~Hz. The axial trap frequency gives the Lamb-Dicke parameter of $\eta~=~0.16 $ for the probing wavelength at 689~nm, which satisfies the Lamb-Dicke condition \cite{Leibfried2003}.

\section{Ramp-up distance calibration} \label{app:Ramp}

To calibrate the ramp-up distance per unit of magnetic field change, we bring the single-frequency red MOT to a lower position than the normal position. Instead of setting ${B}^{\text{bias}}_{\text{z}} = 1.8$ G as usual, we set ${B}^{\text{bias}}_{\text{z}} = 1.9$ G. This allows us to horizontally image atom cloud clearly as they are finally pushed up and loaded into the lattice trap.

\begin{figure}[t!]
     \includegraphics[width=\columnwidth]{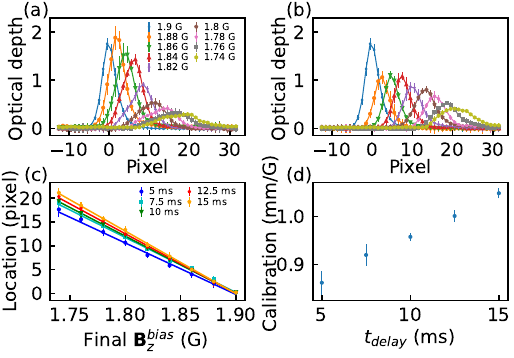}
     \caption {\textbf{Ramp-up distance calibration.} We calibrate the ramp-up distance per unit of bias magnetic field by taking the vertical slices of the optical depth from the absorption images of the lattice atom cloud at various final ${B}^{\text{bias}}_{\text{z}}$ magnitude. Data in (a) is taken at $t_{\text{delay}} = 5$ ms, while (b) is of $t_{\text{delay}} = 15$ ms. As mentioned before, a smaller $t_{\text{delay}}$ yields more atoms finally, but less distance. We fit each optical depth slice to a Gaussian profile to determine the center position of the atom cloud. (c) We plot the atom cloud's vertical location versus the final ${B}^{\text{bias}}_{\text{z}}$ magnitude under different $t_{\text{delay}}$ settings. We fit a linear function to the data. (d) Plot of the slope of the linear fit in (c) versus $t_{\text{delay}}$ setting. The distance is calibrated by using the pixel width of $8\ \mathrm{\mu m}$.   
     }
   \label{fig:ramp_calibration}
\end{figure}

As described in the main text, from the single-frequency red MOT position, the atoms are ramped up by changing the ${B}^{\text{bias}}_{\text{z}}$ for the final time. Then, the lattice beam is abruptly turned on at $t_{\text{delay}}$ after the magnetic field ramp. We take absorption images of the atom cloud at different final values of ${B}^{\text{bias}}_{\text{z}}$. At each field ramp value, we take a vertical slice of the atom cloud average optical depth image. Respectively, Fig. \ref{fig:ramp_calibration}(a) and (b) show the vertical slice of the atom cloud at different final ${B}^{\text{bias}}_{\text{z}}$ for $t_{\text{delay}} = 5$ ms and $t_{\text{delay}} = 15$ ms. We perform this calibration for $t_{\text{delay}} = 5,\ 7.5,\ 10,\ 12.5,\ \text{and}\ 15$ ms.  

For each vertical slice, we fit a Gaussian profile to the data to determine the center vertical position of the atom cloud. For each $t_{\text{delay}}$, the vertical location of the atom displays a linear relationship with the final ${B}^{\text{bias}}_{\text{z}}$ [Fig. \ref{fig:ramp_calibration}(c)]. Here, the data is shown in units of the number of image pixels. For our horizontal imaging system, we use the conversion of $8\ \mathrm{\mu m}$ per pixel. The ramp-up distance per unit of bias magnetic field  calibration is plotted against the corresponding $t_{\text{delay}}$ setting [Fig. \ref{fig:ramp_calibration}(d)]. Interestingly, this data also shows a linear relationship.
\section{Optical lattice characterization} \label{app:LatticeChar} 

We characterize the 1-dimensional magic-wavelength optical lattice trap at the close and far locations by measuring the sideband-resolved absorption spectrum of the trapped atoms. In the Lamb-Dicke regime, ground state atoms in a harmonic trapping potential have three excitation paths to the excited state. The first path is the excitation from a motional state of the ground state ${}^1S_0$ harmonic potential to the same motional state of the excited state ${}^3P_1$. This is the carrier frequency, which is at the atomic transition frequency. The second and third paths are excitations to one lower and higher motional state of the ${}^3P_1$ state. Respectively, these are the cooling sideband and the heating sideband as one is lowering the atom to the ground motional state while the other raises the atom's motional state. The cooling and heating sidebands are red-detuned and blue-detuned from the carrier, respectively. The frequency detuning of the sidebands is equal to the harmonic trapping frequency. Here, only the axial harmonic trapping frequency is resolvable by the $\approx 15$ kHz linewidth spectroscopy profile.   

In our experiment, lattice sidebands are absent in the fluorescent spectrum. When probing atoms at the cooling sideband, they are quickly cooled to the ground motional state, rendering them unable to further absorb the photons. Conversely, when probing the heating sideband, atoms are removed from the trap before a sufficient number of photons are captured. Meanwhile, probing at the carrier frequency allows the excitation and emission cycle to happen over a much longer time before the atoms are lost. 

\begin{figure}[t!]
     \includegraphics[width=\columnwidth]{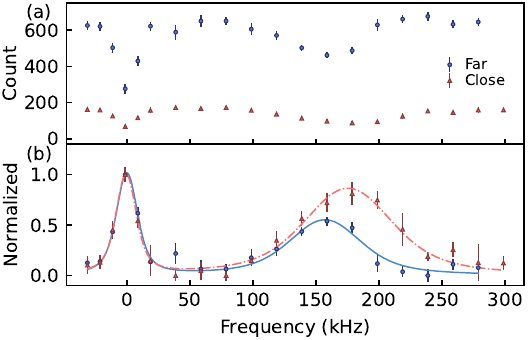}
     \caption{\textbf{Resolved sideband spectroscopy.} (a) Raw data of resolved sideband spectroscopy of the optical lattice at the far and close lattice location. The dips in the photon count show the frequency detunings at which the heating pulse removes atoms from the lattice trap. This heating process is the most efficient on resonance and at the heating sideband frequency. At the cooling sideband, atoms are cooled to the bottom of the trapping potential, resulting in no loss of atoms. (b) The data is normalized against the maximum and the minimum photon counts of the scan. We fit this data to determine the sideband frequency separation from the resonance carrier frequency.}
   \label{fig:lattice_data}
\end{figure}

Our sideband spectroscopy technique relies on measuring the number of atoms remaining in the trap after being exposed to a heating light pulse. The heating light pulse is from the 689-nm probe beam, whose frequency is stepping from the carrier to the heating sideband. The heating pulse is $50$ ms long. After the heating pulse, the 689-nm probe beam's frequency is switched to the carrier resonance frequency, where a fluorescent spectroscopy sequence is performed. If the heating pulse was at the carrier frequency or the heating sideband, the number of photons recorded is significantly reduced, indicating atom loss (Fig. \ref{fig:lattice_data}). 

We fit the normalized data to the double Voigt profile. At the far position, the sideband detuning is $157(3)$~kHz. For the close location, it is $176(4)$~kHz. Given the approximate power of the trapping beam, the measured reflectance of the test surface, and the polarizability of the atom, we calculate that the beam radii at the far and close locations are $24\ \mathrm{\mu m}$ and $22\ \mathrm{\mu m}$ respectively. Clearly, the beam is focused onto the surface.

\section{Strontium adsorption to the surface} \label{app:SrAdsorption}
\begin{figure}
  \includegraphics[width=\columnwidth]{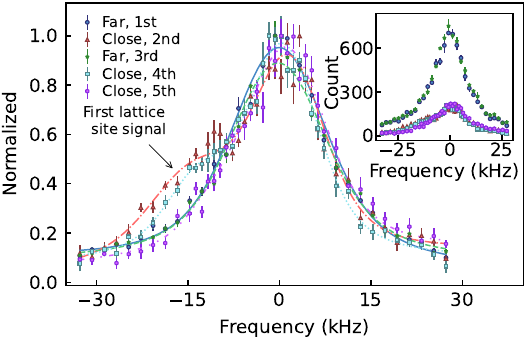}
  \caption{\textbf{Additional spectroscopic measurement data of the CP force.} Normalized spectroscopy data of all five scans at the far (blue circles and green stars) and close (red triangles, cyan squares, and purple hexagons) locations showing the effect on the peaks in response to the strontium coating of the surface. The principal peak is unaffected, while the secondary peak due to the CP potential shift is slowly disappeared. \textbf{Inset}: Raw photon count data of the scans.}
   \label{fig:CPdatSupp}
\end{figure}

We monitor the fluorescent spectrum profiles as we continue bringing the atoms to the surface by doing three extra scans. That is a total of five scans at the far and close locations. In general, the first and the second scans, at far and close locations respectively, are measurements with a clean surface. Then, the third scan is repeated at the far location, showing that the principal resonance peak remains unaffected. The fourth and fifth scans are used to monitor the secondary peak.

Interestingly, the fourth scan shows the secondary peak height reduced, where the fitted curve gives a frequency shift of $-14.1\pm0.9$ kHz from the principal peak. The fifth scan is done again at the close distance, and it shows no sign of a secondary peak. We attribute this effect to strontium atoms adsorbed to the test surface. 

This speculation is further confirmed by observing that the secondary peak reappears when we align the optical lattice beam to focus onto different locations of the surface. We often need to move the lattice over by roughly $40\ \mathrm{\mu m}$, about the lattice beam waist diameter. This suggests that the strontium atoms coating the test surface are brought up predominantly by the lattice beam. We estimate that the process of strontium adsorption to the surface occurs over a few hundred experimental cycles when the atoms are loaded into the close lattice site. Each close spectroscopy scan is performed over $336$ experimental cycles.

\end{appendix}

\end{document}